\documentclass[a4paper, amsfonts, amssymb, amsmath, reprint, showkeys, nofootinbib]{revtex4-1}
\usepackage[english]{babel}
\usepackage[utf8]{inputenc}
\usepackage[colorinlistoftodos, color=green!40, prependcaption]{todonotes}
\usepackage{amsthm}
\usepackage{mathtools}
\usepackage{physics}
\usepackage{xcolor}
\usepackage{graphicx}
\usepackage[left=23mm,right=13mm,top=35mm,columnsep=15pt]{geometry} 
\usepackage{adjustbox}
\usepackage{placeins}
\usepackage[T1]{fontenc}
\usepackage{comment}

\usepackage[hidelinks]{hyperref}
\begin{document}
\title{Precision timing at the HL-LHC with the CMS MIP Timing Detector: current progress on validation and production}

\author{\textit{Presented at the 32nd International Symposium on Lepton Photon Interactions at High Energies, \\ Madison, Wisconsin, USA, August 25–29, 2025} \\ \vspace{0.5cm} Simona Palluotto on behalf of the CMS Collaboration}
    \email[Correspondence email address: ]{s.palluotto@campus.unimib.it}
    \affiliation{INFN \& Università degli Studi di Milano-Bicocca, Milano, Italy}

\begin{abstract}
During the High Luminosity phase of LHC, up to 200 proton-proton collisions per bunch crossing will bring severe challenges for event reconstruction. To mitigate pileup effects, an extended upgrade program of the CMS experiment is expected. A new timing layer, the MIP Timing Detector (MTD), will be integrated between the tracker and the calorimeters. With a time resolution of 30-60 ps, the MTD will enable 4D vertexing and it will bring significant improvements in track-to-vertex association and object identification. The MTD is composed of two subsystems based on different technologies: the Barrel Timing Layer (BTL) consists of LYSO:Ce scintillating crystals readout by SiPMs and the Endcap Timing Layer (ETL) is made of Low Gain Avalanche Detectors. The BTL is currently under production, while ETL sensor prototyping and validation are ongoing. Recent system tests have confirmed the performance of the full acquisition chain. This talk will provide an overview of the MTD design, along with the physics motivation and the current status of BTL construction and ETL development.
\end{abstract}


\maketitle

\section{Introduction} \label{sec:outline}

The LHC has demonstrated its dual role as both a discovery machine and a precision instrument. Large datasets, together with advanced reconstruction techniques, have enabled measurements of high accuracy. To achieve the precision necessary for studying very rare processes, testing further the Standard Model and exploring potential new physics, a higher-luminosity phase is required. The High-Luminosity LHC (HL-LHC), expected to begin operation in 2030, will provide an increase in luminosity, delivering proton-proton collisions at 14\,TeV with integrated luminosities of 3000\,fb$^{-1}$ for experiments such as CMS and ATLAS~\cite{HL_LHC}.
However, operating under these extreme conditions poses major challenges, requiring detectors capable of withstanding high radiation and particle rates.
To maintain performance under extreme pileup, rising from the current $\mathcal{O}(50)$ up to $\mathcal{O}(200)$, CMS will integrate a new timing detector capable of measuring the arrival time of charged particles with a precision of about 30\,ps at the start of HL-LHC operation, degrading to around 60\,ps by the end of operation. The MIP Timing Detector (MTD)~\cite{MTD_TDR} covers both the barrel (BTL) and endcap (ETL) regions up to a pseudorapidity of $|\eta|=3$, using technologies optimized for their respective environments. \\

\begin{figure*}[!htbp] 
    \centering
    \includegraphics[height=6.5cm]{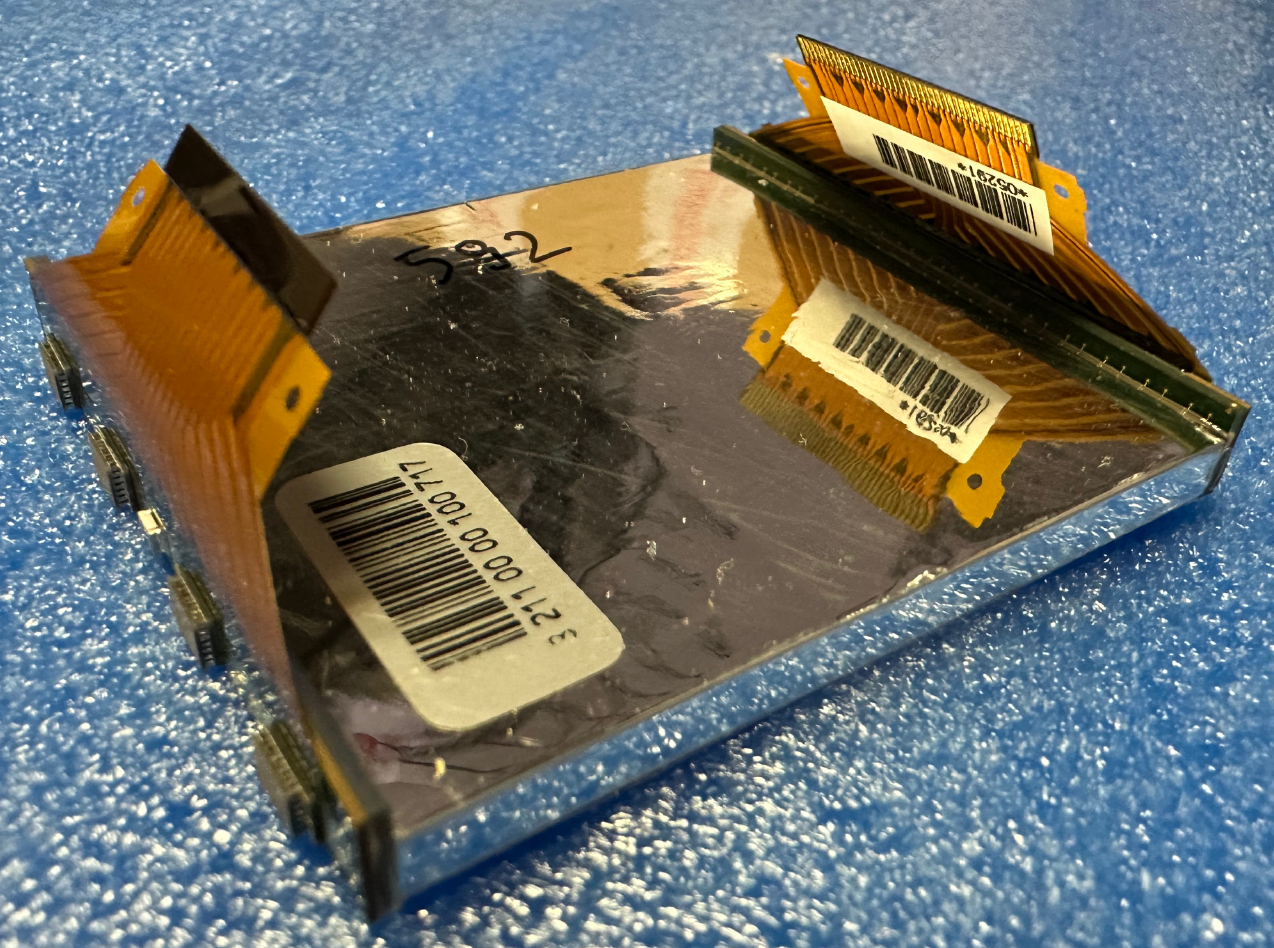}
    \includegraphics[height=6.5cm]{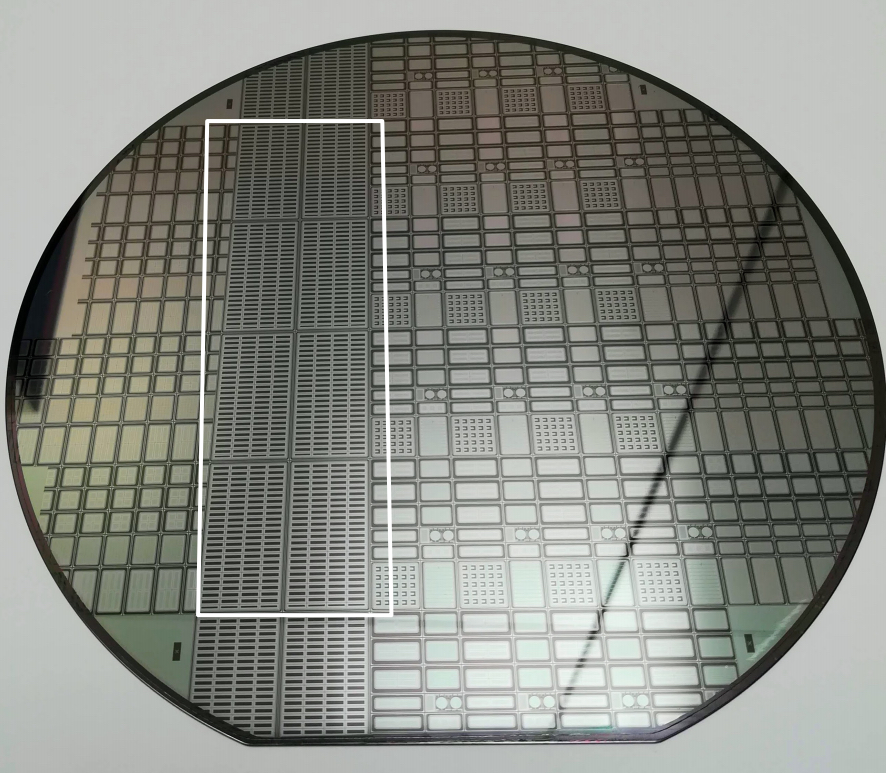}
    \caption{The sensor technologies selected for the barrel and endcap regions of the MTD are shown on the left and right, respectively. The BTL sensor module, displayed on the left, consists of an array of 16 LYSO:Ce scintillating bars, each read out at both ends by SiPMs with an active area matching the crystal bar end face. On the right, a wafer of LGADs is shown, representing the sensor technology that will instrument the ETL.}
    \label{fig:mtd_pieces}
\end{figure*}

Comprehensive R\&D and production protocols have been established to ensure the target performance. Mass production of the BTL is currently underway, with installation expected by the end of 2026, while ETL production and installation are planned for completion by mid-2029. The two sub-systems employ different technologies to meet their requirements in radiation tolerance, cost, schedule, and mechanical constraints. In particular, the ETL inner radius experiences approximately 30 times the fluence of the BTL, while the BTL surface is about 2.5 times larger. Mechanically, the BTL must operate maintenance-free within the tracker cold volume, whereas the ETL is housed in an independently cooled and periodically accessible volume. These differing constraints determined the technology choice: the BTL uses LYSO:Ce crystal bars read out at both ends by Silicon Photomultipliers (SiPMs), whereas the ETL employs Low Gain Avalanche Diodes (LGADs). The two technologies are shown in Figure~\ref{fig:mtd_pieces}.\\

Precise timing information will enhance vertex and track reconstruction, as well as pileup rejection, by exploiting the fact that collision vertices have a time spread of $\mathcal{O}$(200\,ps). Achieving a time resolution of tens of picoseconds allows overlapping vertices in space to be separated in time~\cite{MTD_vertex}, enabling the transition from 3D to 4D vertex reconstruction and therefore reducing pileup contamination. This improvement leads to better object reconstruction, including MET, b-tagging, and lepton isolation efficiencies, ultimately increasing the physics sensitivity of many analyses, particularly those with multi-object final states such as di-Higgs searches, where a gain equivalent to 2--3 additional years of data taking is expected. 
The MTD will also provide new capabilities in CMS, including time-of-flight measurements, which can improve particle identification and enhance sensitivity in various BSM searches, such as those for long-lived particles~\cite{MTD_physics}. \\

\section{The Barrel Timing Layer} \label{sec:btl}
The BTL will instrument the CMS detector up to $|\eta|=1.48$, covering a total surface of 38\,m$^2$ with a thickness of about 4\,cm. It will be installed between the tracker and the electromagnetic calorimeter and will consist of LYSO:Ce crystal bars read out at both ends by SiPMs, for a total of 331,776 readout channels. Each sensor module (SM) comprises an array of 16 LYSO:Ce bars coupled to SiPM arrays. 
The sensor technologies have been chosen thanks to several particularly favorable characteristics.
LYSO:Ce crystals offer high radiation tolerance, a large light yield of about 40,000\,ph/MeV, a fast scintillation rise time ($\tau < 100$\,ps), and a short decay time ($\tau \sim 40$\,ns). The bar-like geometry minimizes the required SiPM area relative to the crystal volume, with typical bar dimensions of $3.12 \times 3.75 \times 54.7$\,mm$^3$. The SiPMs are compact, radiation-tolerant, and insensitive to magnetic fields, with an active area matching the crystal bar end face.\\

\begin{figure*}[!htbp] 
    \centering
    \includegraphics[height=6.8cm]{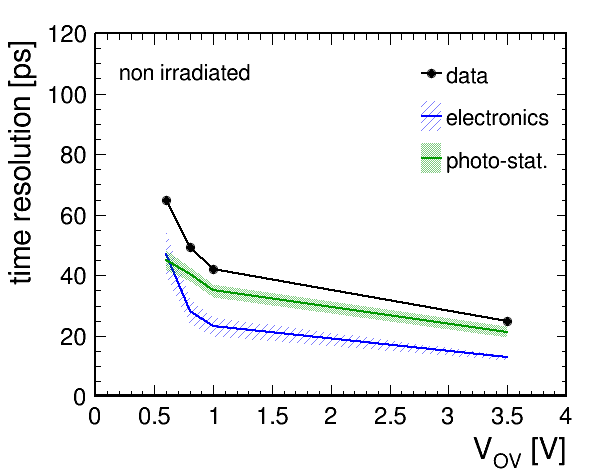}
    \includegraphics[height=6.8cm]{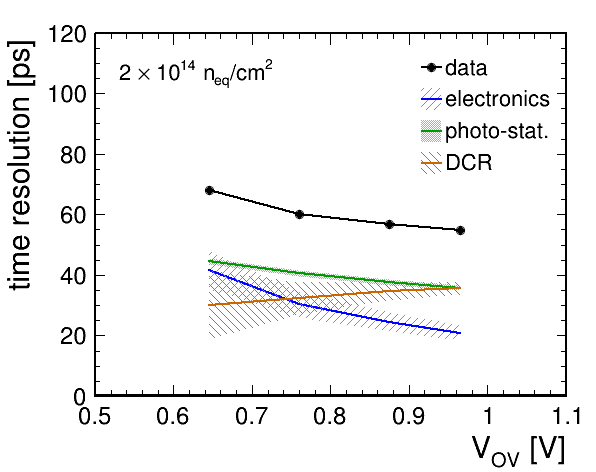}
    \caption{The results of the test beam measurements are shown with black markers, representing non-irradiated modules on the left and modules irradiated to the full fluence expected at the end of operation on the right. The contributions to the overall performance are indicated by colored bands, with photo-statistics, electronic noise, and DCR represented in green, blue, and orange, respectively. Uncertainties are included in the plot as error bars on the points and as bands for the individual contributing terms~\cite{btl_confirmation}.}
    \label{fig:btl_performance}
\end{figure*}

The time information is extracted independently from both ends of each crystal bar, providing two nearly uncorrelated time measurements for a particle traversing the bar. The overall time resolution of the BTL can be expressed as:
\begin{equation}
    \sigma_{t}^{\mathrm{BTL}} \sim \sigma_t^{\mathrm{noise}} \oplus \sigma_t^{\mathrm{photo-statistics}} \oplus \sigma_t^{\mathrm{DCR}} ,
\label{eq:btl_tres}
\end{equation}
where $\sigma_t^{\mathrm{noise}}$ represents the contribution from electronic noise, $\sigma_t^{\mathrm{photo-statistics}}$ arises from fluctuations in photon arrival times at the SiPMs, and $\sigma_t^{\mathrm{DCR}}$ accounts for degradation due to radiation induced dark count rate (DCR),  with the expected fluence reaching up to 1~MeV neutron equivalent fluence, n$_{eq}$/cm$^2$, of $2 \times 10^{14}$\,n$_{eq}$/cm$^2$. The expected performance at the beginning and end of HL-LHC operation is shown in Figure~\ref{fig:btl_performance}, with the main contributions represented by the colored bands. \\

A major challenge during the R\&D phase has been the mitigation of radiation damage, which induces a DCR level up to 20\,GHz per SiPM over the detector lifetime. Several complementary strategies have been adopted to address this issue: reducing the SiPM overvoltage during operation from 3.5\,V to about 1\,V, implementing dedicated noise filtering in the TOFHIR readout ASIC~\cite{tofhir2}, and integrating Thermoelectric Coolers (TECs)~\cite{tecs} on the backside of the SiPM arrays. The TECs enable a tenfold reduction in the SiPM dark count rate by maintaining an operating temperature of $-45^{\circ}$C, with periodic in situ annealing at $60^{\circ}$C to recover SiPM performance. 
In parallel, the light output and the signal slope of the crystal-SiPM system have been optimized through extensive studies of module thickness and SiPM cell size, leading to the choice of 3.75\,mm-thick sensors and 25\,$\mu$m SiPM cells. These parameters maximize the light output performance while maintaining reduced dark noise.\\

\begin{figure*}[!htbp] 
    \centering
    \includegraphics[height=6.8cm]{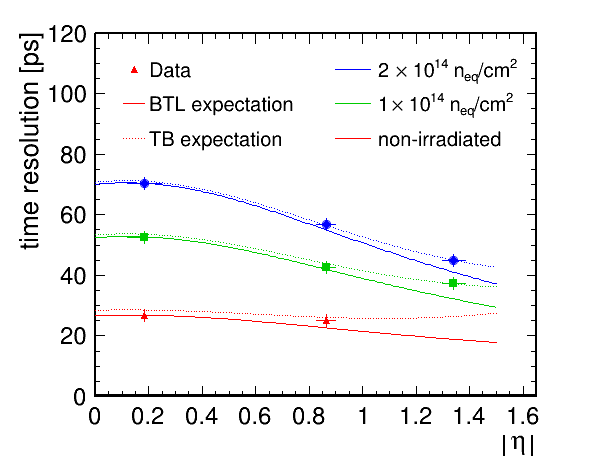}
    \includegraphics[height=6.8cm]{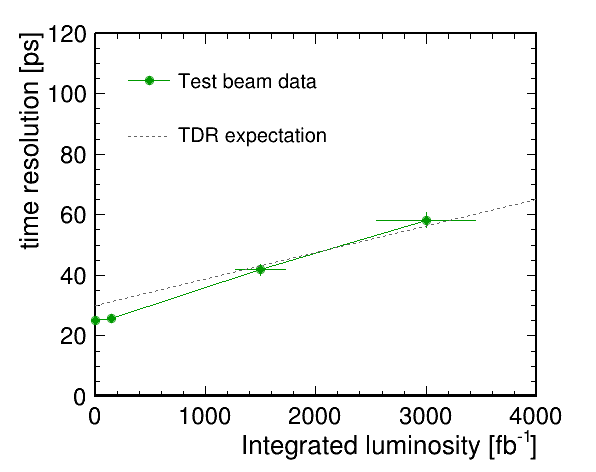}
    \caption{Left: Measured time resolution as a function of the equivalent pseudorapidity for modules exposed to different irradiation levels. Dots represent the data, while solid lines show the BTL model predictions. Dashed lines indicate the expected test beam resolution, which includes an additional contribution from the experimental conditions. Right: Time resolution as a function of the equivalent integrated luminosity. Measurements are shown for modules with non-irradiated SiPMs and for SiPM arrays exposed to different fluences. The dotted line marks the target resolution defined in the TDR~\cite{btl_confirmation}.}
    \label{fig:btl_performance_lumi_eta}
\end{figure*}

The expected BTL performance across the full pseudorapidity range and throughout the HL-LHC lifetime has been validated through test-beam campaigns using sensor modules irradiated to various fluences ($1\times10^{13}$, $1\times10^{14}$, and $2\times10^{14}$\,n$_{\mathrm{eq}}$/cm$^2$). The corresponding DCR levels emulate different operation phases of the detector, and modules were tested at various tilting angles to reproduce the energy deposition at different pseudorapidities. The measured results, shown in Figure~\ref{fig:btl_performance_lumi_eta}, are consistent with design expectations: non-irradiated modules achieve about 25\,ps time resolution, degrading to roughly 55\,ps at the end of operation.\\

These validated performance measurements marked the start of the large-scale assembly of the BTL. Two sensor modules, each equipped with front-end electronics, are combined to form a Detector Module (DM), which is housed in a copper housing for optimal thermal and mechanical stability. Twelve DMs are then mounted on a cold tray to form a Readout Unit (RU), and six RUs constitute a single tray. In total, 72 trays will instrument the entire BTL surface. Production is distributed among four BTL Assembly Centers (BACs) located at Beijing, Caltech, Milano, and the University of Virginia, operating with common assembly and QA/QC procedures. The first tray was successfully assembled in March, marking the start of full-scale production across all BACs. 
BTL production is currently progressing steadily, with 100\% of the LYSO crystals and SiPMs produced and qualified. Approximately 80\% of the sensor modules have been assembled, demonstrating consistent quality across all BACs. More than 50\% of the detector modules have been produced, and about 10\% of trays have already been assembled and tested. The completion of tray production is expected by February~2026.
\section{The Endcap Timing Layer} \label{sec:etl}

The endcap region is instrumented with LGADs, forming the ETL. The ETL will be installed in front of the forward High Granularity Calorimeter and outside the forward tracker volume. It consists of two disks on each side of the detector, covering a total active area of about 14\,m$^2$, and providing two independent time measurements per track. The design target is a time resolution of 35\,ps per track, degrading to less than 50\,ps at the end of the detector lifetime.
The ETL will operate in a challenging radiation environment, with the expected fluence reaching up to $1.6 \times 10^{15}$\,n$_{eq}$/cm$^2$. \\

The time resolution is primarily limited by two contributions: jitter, caused by electronic noise affecting the signal timing, and ionization fluctuations, which vary from event to event due to the stochastic nature of charge creation along the particle path and, for 50\,$\mu$m-thick sensors, it is expected to give a contribution of about 25 ps to the time resolution.\\

To meet the performance requirements, extensive R\&D has been carried out on LGAD technology. The gain layer has been optimized to achieve a gain in the range of 10--30. The optimal sensor thickness has been identified as 50\,$\mu$m, providing a fast signal rise time and reduced Landau fluctuations. ETL modules will employ 16$\times$16 LGAD arrays characterized by low capacitance, and a pitch of 1.3\,mm.
Significant efforts have also been devoted to improving the radiation tolerance of LGAD sensors. The main degradation mechanism, gain reduction due to acceptor removal in the multiplication layer, is mitigated through higher bias voltages and the introduction of carbon co-implantation in regions exposed to higher fluence. However, the bias voltage cannot be increased arbitrarily due to the risk of single-event burnout, i.e. a failure mode caused by rare, high-energy ionizing events that create localized conductive paths, which has been observed only for bulk electric fields exceeding $E_\mathrm{bulk} \geq 11.5$\,V/$\mu$m. Prototype LGADs tested before and after irradiation have demonstrated compliance with ETL performance requirements for $E_\mathrm{bulk} < 11.5$\,V/$\mu$m.\\

\begin{figure*}[!htbp] 
    \centering
    \includegraphics[height=7.5cm]{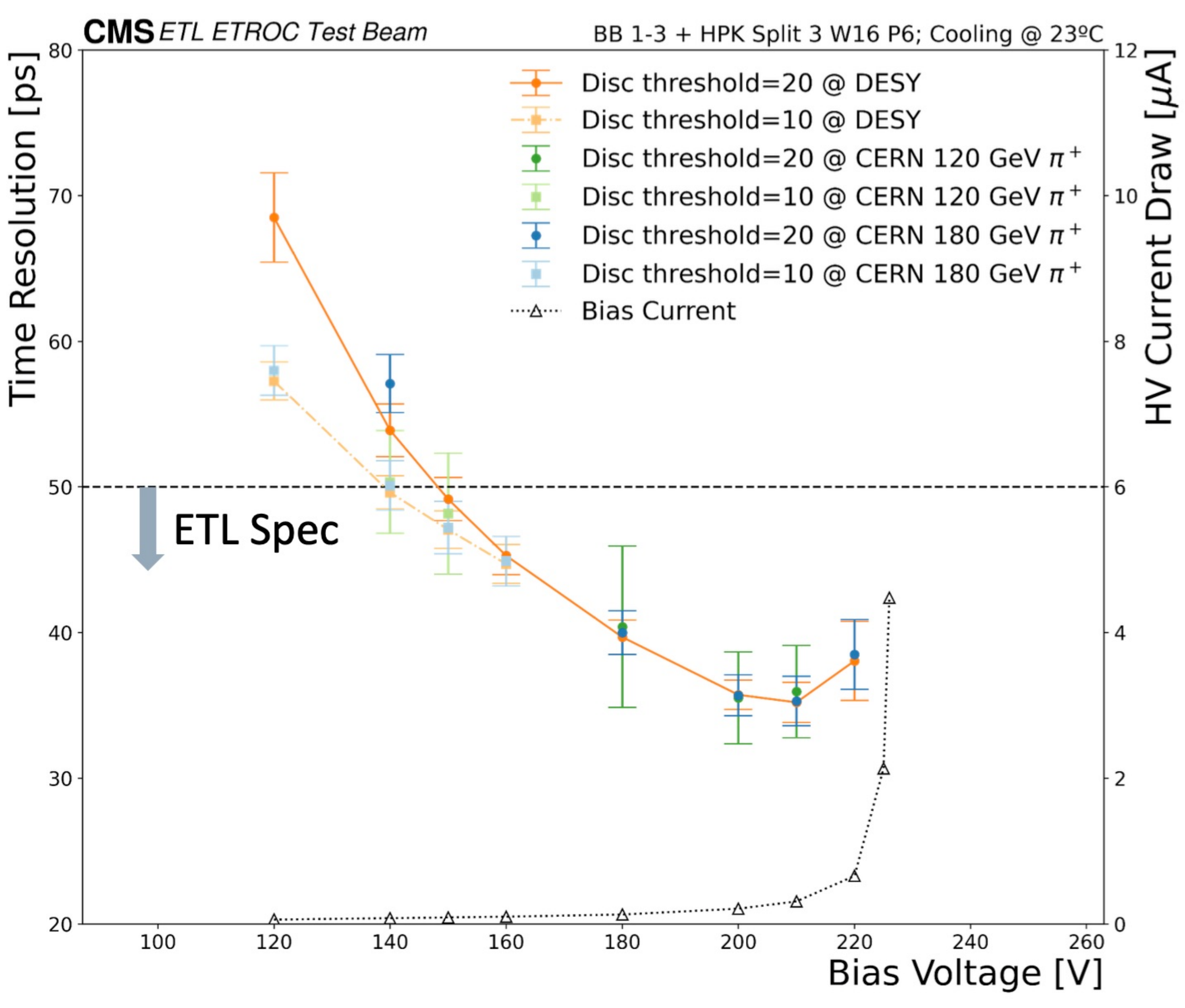}
    \caption{Data from ETROC and large 16$\times$16 arrays of LGADs demonstrate that the target performance is achieved within a bias range of approximately 70\,V~\cite{etl_valentina}.}
    \label{fig:etl_performance}
\end{figure*}

The LGAD sensors are read out by the Endcap Timing Read-Out Chip (ETROC), a custom low-noise, low-power ASIC designed to operate reliably in the harsh radiation environment of the endcap region. Its power consumption is about 1\,W per chip at 1.2\,V and less than 4\,mW per channel. The combined LGAD+ETROC system has undergone extensive validation, with final prototypes achieving the target time resolution across a wide bias range of about 70\,V~\cite{etl_valentina}, as shown in Figure~\ref{fig:etl_performance}.
Each ETL LGAD matrix measures approximately $2 \times 2$\,cm$^2$, and each ETL module consists of four LGAD matrices, one ETROC chip, a baseplate, and a dedicated PCB for interfacing with the readout system. In total, the ETL will include about 8 million readout channels. The production of roughly 8000 ETL modules will be distributed across four assembly sites, i.e. Fermilab, Boston University, INFN (Italy), and IFCA (Spain), where standardized quality assurance and control procedures are being finalized to ensure uniform performance across the entire detector.

\section{Conclusions} \label{sec:conclusions}
The upcoming high luminosity phase of LHC represents a paradigm shift for the CMS experiment, with the integration of a new precision timing detector enabling full four-dimensional event reconstruction. This capability is crucial to mitigate the effects of the extremely high pileup conditions expected at the HL-LHC, while simultaneously providing new measurement opportunities and extending the overall physics reach of CMS. The inclusion of precise timing information is anticipated to significantly enhance the sensitivity of a broad range of analyses and searches. The MTD will provide coverage up to $|\eta| = 3$ and is composed of two subsystems based on distinct technologies. 
The BTL, employing LYSO:Ce crystals read out by SiPMs, has completed validation and fully meets the target design specifications, marking the start of large-scale module production, with completion expected by 2026. The ETL, based on LGADs, has finalized its component design and successfully validated its target performance, with installation foreseen for 2029.

\section*{Acknowledgements} \label{sec:acknowledgements}
Results reported in this work are part of a collaborative effort, and we gratefully acknowledge the support of the funding agencies that contributed to the design and construction of the CMS MTD.

\end{document}